%% file: main.tex
\documentclass[journal,twoside,web]{ieeecolor}
\usepackage{generic}
\usepackage{cite}
\usepackage{amsmath,amssymb,amsfonts}
\usepackage{algorithmic}
\usepackage{graphicx}
\usepackage{textcomp}

\usepackage{bbding}
\usepackage{multirow}
\usepackage{subcaption}
\usepackage{caption}

\begin{document}


\title{Automated Mobility Context Detection with Inertial Signals}
\author{Antonio Bevilacqua, Lisa Alcock, Brian Caulfield, Eran Gazit, Clint
  Hansen, Neil Ireson, Georgiana Ifrim

  \thanks{Antonio Bevilacqua and Georgiana Ifrim are with the Insight Centre for
  Data Analytics, School of Computer Science, University College Dublin,
  Ireland (e-mail: \{antonio.bevilacqua, georgiana.ifrim\}@ucd.ie).}

\thanks{Brian Caulfield is with the Insight Centre for Data Analytics, School
  of Public Health, Physiotherapy and Sports Science, University College
  Dublin, Ireland (e-mail: brian.caulfield@ucd.ie).}  \thanks{Lisa Alcock is
  with the Clinical Ageing Research Unit, Faculty of Medical Sciences,
  Newcastle University (email: lisa.alcock@ncl.ac.uk).}

\thanks{Eran Gazit is with the Center for the Study of Movement, Congnition and
  Mobility, department of Neurology, Tel Aviv Sourasky Medical Center (email:
  erang@tlvmc.gov.il).}

\thanks{Clint Hansen is with the Department of Neurology, Kiel University
  (email: c.hansen@neurologie.uni-kiel.de).}

\thanks{Neil Ireson is with the Organisation, Information and Knowledge Group,
  Department of Computer Science, University of Sheffield (email:
  n.ireson@sheffield.ac.uk).}

}

\maketitle

\input{chapters/0.abstract}


\input{chapters/1.introduction}
\input{chapters/2.related}
\input{chapters/3.datasource}
\input{chapters/4.experiments}
\input{chapters/5.results}
\input{chapters/6.discussion}
\input{chapters/7.conclusions}
\input{chapters/8.acknowledgements}
\input{chapters/9.appendix}

\bibliographystyle{plain}
\bibliography{main}

\end{document}

%% file: chapters/0.abstract.tex
\begin{abstract}
  Remote monitoring of motor functions is a powerful approach for health
  assessment, especially among the elderly population or among subjects
  affected by pathologies that negatively impact their walking
  capabilities. This is further supported by the continuous development of
  wearable sensor devices, which are getting progressively smaller, cheaper,
  and more energy efficient. The external environment and mobility context have
  an impact on walking performance, hence one of the biggest challenges when
  remotely analysing gait episodes is the ability to detect the context within
  which those episodes occurred. The primary goal of this paper is the
  investigation of context detection for remote monitoring of daily motor
  functions. We aim to understand whether inertial signals sampled with
  wearable accelerometers, provide reliable information to classify
  gait-related activities as either indoor or outdoor. We explore two different
  approaches to this task: (1) using gait descriptors and features extracted
  from the input inertial signals sampled during walking episodes, together
  with classic machine learning algorithms, and (2) treating the input inertial
  signals as time series data and leveraging end-to-end state-of-the-art time
  series classifiers. We directly compare the two approaches through a set of
  experiments based on data collected from 9 healthy individuals. Our results
  indicate that the indoor/outdoor context can be successfully derived from
  inertial data streams. We also observe that time series classification models
  achieve better accuracy than any other feature-based models, while preserving
  efficiency and ease of use.
\end{abstract}

%% file: chapters/1.introduction.tex
\section{Introduction}
\label{section:intro}

The widespread availability of inexpensive wearable sensor technologies means
that measurement of motor function, particularly gait, on an ongoing basis in
the home and community setting is easier than ever before \cite{pmid29730488},
\cite{s19081925}. This offers enormous potential in the continuous assessment
of gait as people move throughout their daily lives, providing accurate and
objective measurements of real-world motor performance and the deep insights
into health status that this brings.

However, this opportunity comes with some significant challenges, one of which
is the issue of being able to identify the context within which walking
occurred. Previous research has demonstrated that gait performance changes in
different contexts, even in young healthy adults
\cite{10.1371/journal.pone.0228682}, \cite{6944249}. This has noteworthy
implications for the measurement and interpretation of real-world walking in
clinical populations \cite{s21030813}. If we can identify the context for
walking performance, it can provide insights into the specific intrinsic and
extrinsic factors that can adversely affect gait in clinical populations,
leading to informed interventions. For example, identifying the extrinsic
factors that lead to deterioration of gait in the elderly population can lead to
recommendations for specific remedial strategies to reduce risk of falling
\cite{pmid26737531}. It is clear, then, that understanding context is critical
in our interpretation of data related to real-world walking. The challenge in
practical implementation is that of identifying context in an unobtrusive manner,
without impacting on the privacy of the individual, or their quality of life.
Different approaches to context identification have been implemented in the
past. Smeaton \textit{et al.} \cite{DBLP:journals/jaise/SmeatonLC12} used small
wearable cameras to capture images that were subsequently used as the basis of
a context annotation model. This was successful from a technical perspective
but has obvious disadvantages from a data processing and privacy standpoint, to
the point that it is not a feasible option. Other approaches have involved use
of data from GPS sensors \cite{pmid25390297}, which also suffers from drawbacks
such as battery duration, privacy management, and poor indoor availability.

In this context, wearable sensors, or Inertial Measurement Units (IMU), offer
a number of advantages when compared with other technologies. They are
inexpensive, built for ease-of-use, and bear a significantly
reduced impact on the users' privacy. They measure physical quantities such as the body's  
acceleration and angular velocity, that come in the form of streams of
continuous values sampled at a fixed frequency \cite{GODFREY20081364}. The
core challenge we face, then, is to understand whether it is possible to
extract mobility context only relying on IMU signals.
The most straightforward approach to this task involves the extraction of a
number of descriptors from the sampled inertial signals. These features, often
called Digital Mobility Outcomes (DMO), are highly specialized for gait-related
activities, and can be used to train a variety of learning models. Examples of
DMOs are the number of steps, the length and duration of a walking episode, and
the gait asymmetry. DMOs can be used for different tasks related to human
movement analysis, including distinction between recreational and utilitarian
walking \cite{kang2017differences} and identification of pathological gait
patterns\cite{Polhemuse038704}. Whilst DMOs are easy to interpret, algorithms
that extract them often require fine tuning and meticulous scrutiny in order
to avoid cherry-picking and ensure their generalization capability.

A different strategy for IMU context identification is based on more generic
Time Series Classification (TSC) methods. As the inertial signals obtained from
IMUs are sequential values sampled at a constant rate, they constitute a set of
time series. This type of data is ubiquitous and, therefore, studied across many
research fields \cite{DBLP:journals/corr/BagnallBLL16}. TSC models do not
require any manual feature engineering steps, and can easily adapt to different
applications and contexts. The goal of this study is then to explore the task
of automated context detection from digital mobility data, that is, to
understand whether contextual information can be extracted  directly from inertial
signals without the need for geographical localization. Specifically, we aim
to answer the following two questions:

\begin{enumerate}
    \item Is it possible to extract indoor/outdoor contextual information from
    inertial streams sampled during gait-related activity, without the use of
    GPS data?
    \item How do feature-based approaches compare with TSC models in the task of
    indoor/outdoor context detection?
\end{enumerate}

In this study, we analyse some of the recent advances in the development of
highly accurate and scalable machine learning algorithms for TSC and focus on
their application to context detection from IMU signals. We propose a new
methodology for context detection from IMU data, and empirically evaluate
different modeling approaches and algorithms for this task. In particular, we
compare classifiers that only use the raw IMU signal, to those that use
extracted domain-specific features (i.e., DMOs).  We study different modeling
approaches for working with IMU data at different levels of granularity, such
as generic, unrestricted signal windows, or specific gait episodes and walking
bouts. Ultimately, we show that TSC methods achieve better performance than
feature-based models in terms of classification accuracy, and also prove to be
more flexible from an application standpoint as they do not rely on static,
shallow feature sets that may not necessarily be available throughout the full
length of an inertial stream.

The rest of this paper is organized as follows: Section \ref{section:related}
introduces the topic of context detection for gait activity and provides a broad
overview of the most common TSC algorithms. Section \ref{section:datasource}
contains a detailed description of the data collected and used for this study,
as well as the processing pipeline required to produce datasets, annotations,
and class labels. A description of our empirical study is provided in Section
\ref{section:experimentalcampaigns}, while Section \ref{section:results} and Section
\ref{section:discussion} provide a detailed analysis and discussion of results,
respectively, alongside a discussion on limitations of this study. Final remarks
are given in Section \ref{section:conclusions}.

%% file: chapters/2.related.tex
\section{Related Work}
\label{section:related}

\subsection{Digital Mobility and Context Detection}

A DMO is an objective quantity measurable through specific processing of
inertial signals, that can be used as a proxy descriptor for gait-related
physical activities \cite{kluge2021consensus}. DMOs represent useful features
to characterize walking and assess mobility performance. In reference \cite{s20247015},
physical activities of daily-living are monitored in a group of participants
affected by Parkinson's Disease using tri-axial accelerometers, and
corresponding DMOs are computed. Reported results show how reduced mobility relates to an increased severity of motor symptoms, which in turn reflects on
the measured DMOs. Most importantly, self-reported mobility metrics oftentimes
do not match with digitally assessed mobility outcomes. In a different study
\cite{kang2017differences}, a small set of DMOs is used to classify walking
episodes as either recreational or utilitarian. The authors show how duration,
intensity, speed, built environment characteristics and location, and time
distributions of walking bouts, can be sufficient to determine the purpose of a
walking episode. Experimental results are detailed for 651 participants, each
one providing data collected with a single uni-axial IMU over a period of 7
days.

The concept of DMO is, however, still relatively young in the  literature, and just recently DMOs started being adopted in clinical trials as a tool for remote mobility monitoring. In particular, the Mobilise-D consortium strives
to develop protocols for validation and approval of digital biomarkers with
clinical potential among cohorts of patients affected by Parkinson's disease,
multiple sclerosis, chronic obstructive pulmonary disease, and recovering 
from proximal femoral fracture \cite{Rochester2020}. The authors outline a
structured roadmap whose ultimate goal is to design healthcare standards for
DMO adoption in mobility assessment. In this regard, a reviewing protocol is
suggested in reference \cite{Polhemuse038704}.

In more recent work, Del Din \textit{et al.} \cite{del2021body} offer a
systematic review of studies concerned with Body-Worn Sensors (BWS) used in
conjunction with DMOs to detect and quantify Parkinson's disease motor symptoms
and dysfunctions such as tremors, bradykinesia, dyskinesia, postural
instability and others. The authors show how DMOs are regarded to be an
effective and powerful tool for the remote monitoring of physical activities
and performance. Nonetheless, most DMOs lack technical and clinical validity,
as the corresponding algorithms have often been designed and tested only in
controlled environments. Moreover, it is still not clear how well
DMO-extracting techniques scale to unseen clinical cohorts, new sampling
hardware or unexpected operational conditions. Recent work on time series
classification has shown that models relying only on raw signals significantly
outperform feature-based approaches \cite{dempster_etal_2020_rocket}, thus in the
next section we give a general introduction of the state of the art for time series
classification.

\subsection{Time Series Classification}
Research on time series classification develops rapidly and a number of
effective classification methods have been proposed recently 
\cite{DBLP:journals/corr/BagnallBLL16}. Traditionally, distance-based
classification using nearest neighbours is considered a strong baseline
strategy. \emph{Dynamic Time Warping} \cite{ding_etal_2008} (DTW) is an elastic
distance measure  proposed as an alternative to the Euclidean distance, that is
capable of detecting spatial and temporal distortions in time series and yields
robust classification results when used to train 1-NN models \cite{wang2010}. A
different approach is the transposition of time series into more compressed and
interpretable feature spaces. Catch-22 \cite{DBLP:journals/corr/abs-1901-10200}
is a proposed set of 22 heterogeneous features selected from the wider group of
4791 hctsa features \cite{FULCHER2017527}, that results in a classification
accuracy drop of 7\% for a gain in speed, on a group of 93 evaluated datasets.

Another family of classification models, relies on symbolic representation and
approximation of time series. Algorithms such as SAX \cite{10.1145/882082.882086}
and SFA \cite{10.1145/2247596.2247656} can be used to compress continuous time
series into discrete and more compact alphabets. This, in turn, allows for an
easier identification of recurring patterns and words. WEASEL
\cite{DBLP:journals/corr/SchaferL17} leverages multi-resolution SFA approximation
to select statistically significant features, used in a second stage to train a
linear model. Similarly, MrSEQL \cite{DBLP:journals/corr/abs-2006-01667}
generates a wide variety of features using both SAX and SFA, and then uses the
SEQL algorithm \cite{10.1145/2020408.2020519} to efficiently select the best
performing features. MrSEQL allows to train a single linear model, or an ensemble
of models.

More recently, ROCKET \cite{dempster_etal_2020_rocket}, a novel kernel-based
classification method, set the new state-of-the-art for non-ensemble TSC.
ROCKET produces a number of features based on the convolutional product between
the target time series and random shapelets, or \emph{kernels}. The generated
features are then used to train a linear model. ROCKET's latest extension, the
MiniROCKET algorithm\cite{dempster_etal_2020_minirocket}, further improved the
feature generation process, by drastically reducing the kernel space. MiniROCKET
is currently the fastest non-ensemble classification method for time series
classification, with accuracy comparable to many other existing and more complex
methods.

%% file: chapters/3.datasource.tex
\section{Data Sources and Aggregation}
\label{section:datasource}

The Mobilise-D project relies on data collected by a consortium of 5
universities and research centres. Each centre is responsible for recruiting
participants across different clinical cohorts, and collecting heterogeneous
datasets in different scenarios \cite{Rochester2020}.

For the data collection phase, participants were instructed to wear a triaxial
IMU for a period of time of around 2.5 hours. The IMU, worn on the lower back
and kept in place by a neoprene sleeve, samples acceleration and angular
velocity of the subject's body with a frequency of 100 Hz. During the same time
period, a mobile application is responsible for recording the GPS coordinates
of the participant's location.  The application is installed on the
participant's mobile device, and synced with the IMU, prior to the beginning of
data collection. During the 2.5 observation hours, participants are in a
free-living environment, however, they are suggested a set of indoor and
outdoor activities that may be integrated in their regular routine. Proposed
activities are selected among common daily tasks, such as \textit{go to the
  kitchen and get a beverage}, or \textit{go outside and take a short stroll in
  front of your house}.

\subsection{Contextual Labeling}\label{subsection:staypoints}
The raw GPS coordinates collected during the 2.5 hour observation period need
to be aggregated into high-level ground truth labels that characterize the
subjects' activity context. A multi-tier processing framework based on
OpenStreetMap \cite{10.1109/MPRV.2008.80} (OSM) and a Solr database, is
responsible for producing \textit{staypoints}, that are, locations where
subjects spend an extended amount of time without changing their position
significantly. Staypoints are detected by either accounting for the absence of
GPS data, that usually indicates that the phone does not attempt to acquire
location updates as GPS coordinates are deemed stationary, or applying distance
and time thresholds to groups of proximate consecutive GPS coordinates, hence
defining candidate staypoints as centroids of spacial and temporal
clusters. Activities performed sufficiently close to a staypoint are labelled
as indoor, whilst activities performed away from any staypoint are defined as
outdoor. The result of aggregating GPS coordinates is then a set of binary
labels, produced with a frequency of 1 Hz, that describe the context of the
corresponding IMU signal. Each label represents the continuous probability of
that particular activity being performed indoor (value of 1) or outdoor (value
of 0).

\subsection{Dataset Generation}\label{subsection:datasets}

\begin{figure*}[h!]
    \centering
    \includegraphics[width=.85\textwidth]{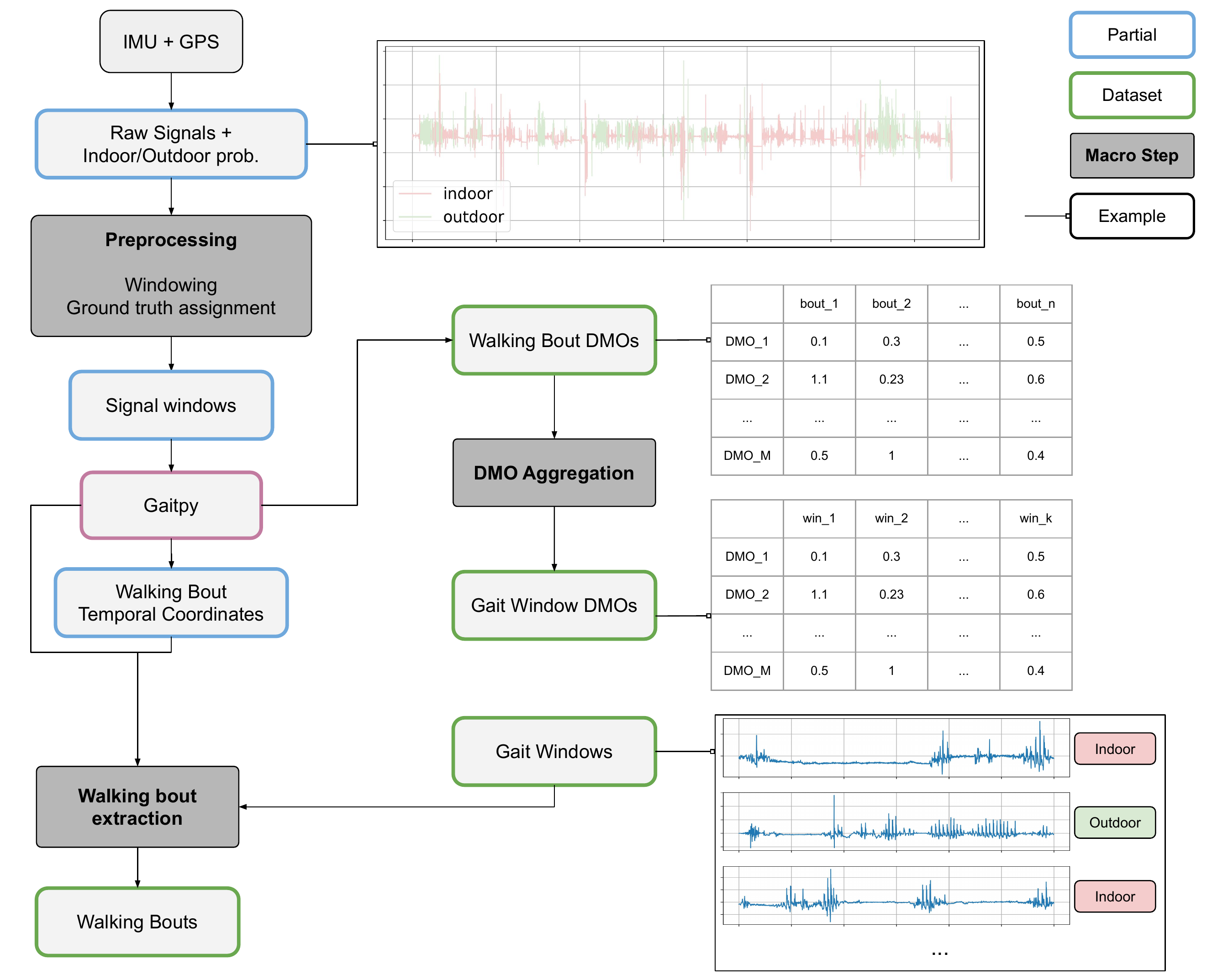}
    \caption{\label{pics:dataflow}Workflow representation of the dataset
      extraction process. Using raw inertial data annotated with GPS-based
      indoor/outdoor labels, 4 datasets can be produced (green boxes): tabular
      features for both walking bouts and gait windows, and inertial signals
      for both walking bouts and gait windows.}
\end{figure*}

The process of generating the datasets from the raw inertial signals is broadly
depicted in Figure \ref{pics:dataflow}. The inertial signals sampled from the
IMU are chunked into small, non-overlapping windows of fixed length. The length
of the windows can be globally adjusted based on the desired model
granularity. In this work we use 1 minute windows. The location probability
values associated with each window are then isolated and aggregated, so that a
unique ground truth class label is produced. The aggregation policy of the
contextual probability values is based on the occurrence of the most frequent
probability value across all the probabilities within a window (i.e., we use a type of majority voting to extract the indoor/outdoor class label of a given window).

\begin{figure}
     \centering
     \begin{subfigure}[b]{\columnwidth}
         \centering
         \includegraphics[width=\columnwidth]{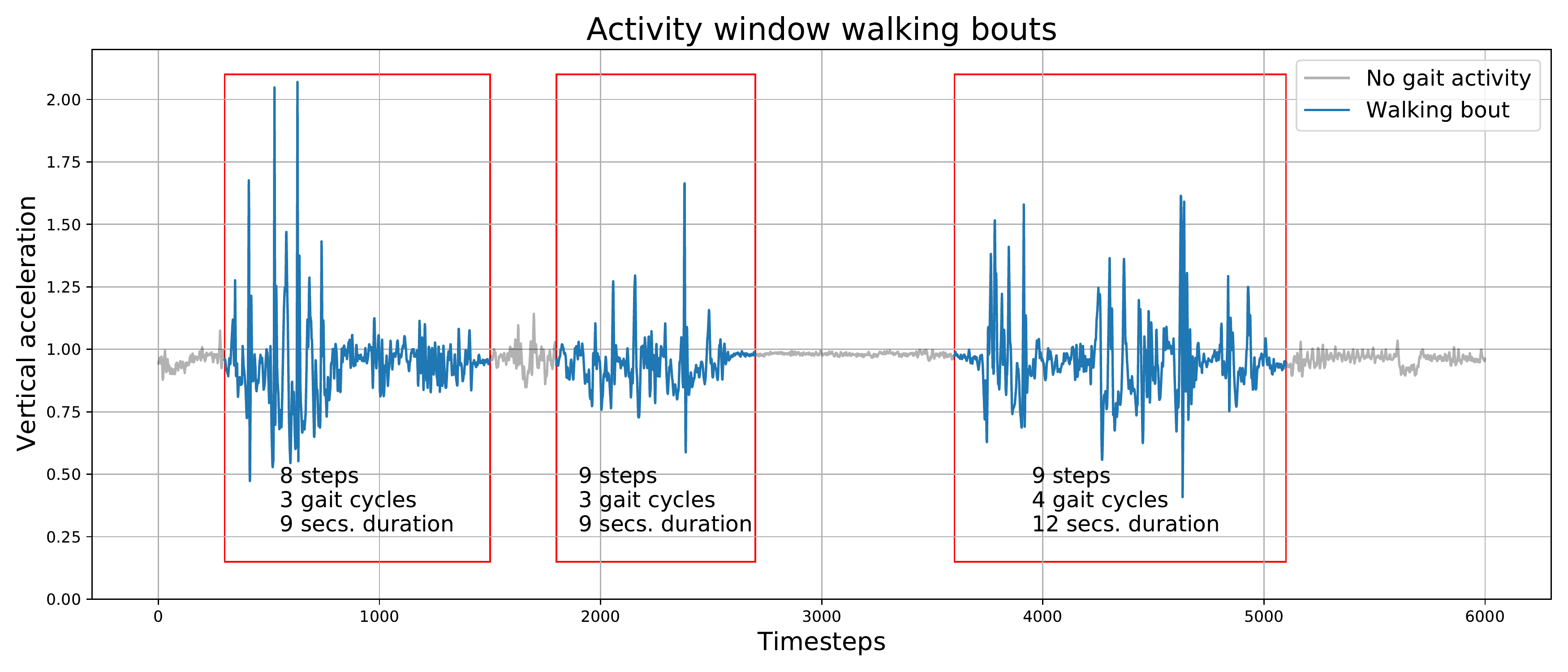}
         \caption{The gait windows dataset contains aggregated features for each
           window, regardless of the number of distinct walking bouts.}
         \label{fig:window}
     \end{subfigure}
     \begin{subfigure}[b]{\columnwidth}
         \centering
         \includegraphics[width=\columnwidth]{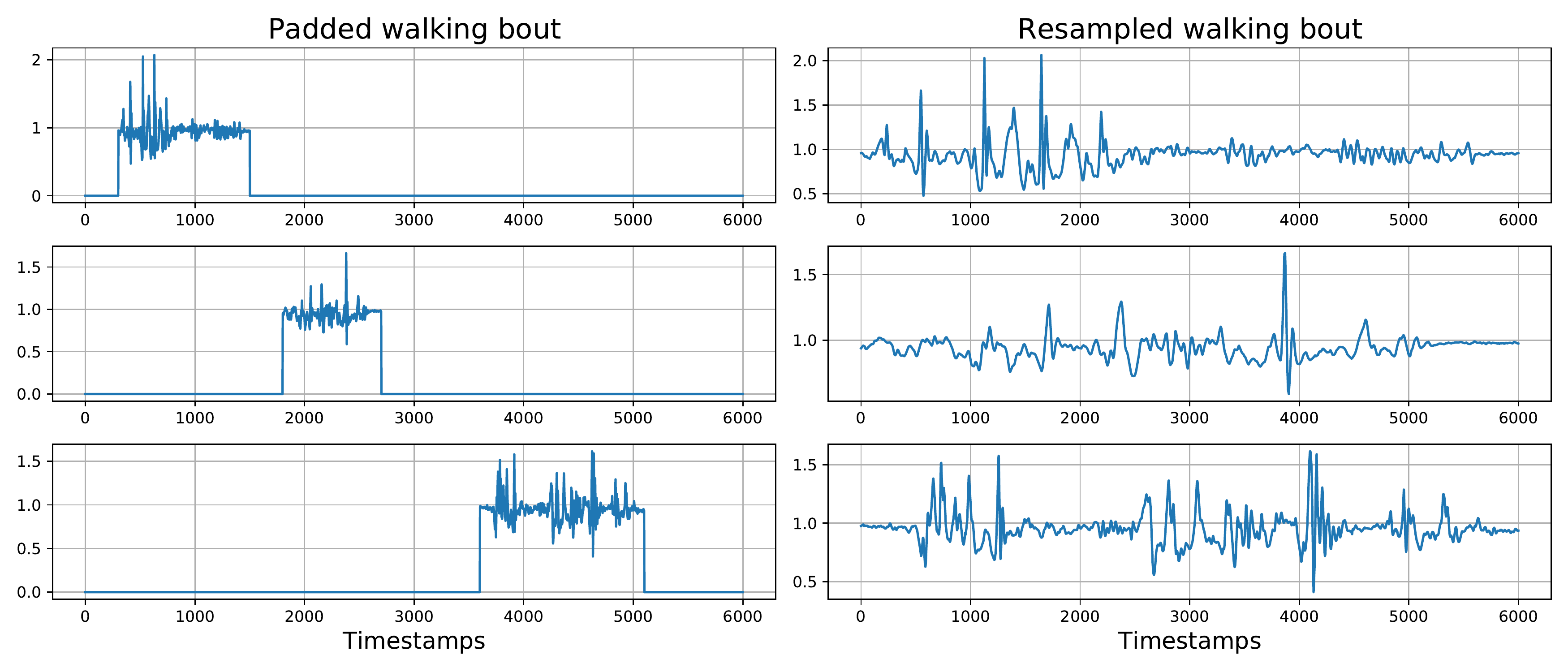}
         \caption{When extracting the walking bout dataset, individual bouts can
           be padded (left) or resampled (right) to match the length of the
           longest possible walking bout, corresponding to 60 seconds.}
         \label{fig:bouts}
     \end{subfigure}
        \caption{Example walking bouts identified by Gaitpy. Once isolated, they can be
          extracted from their original windows and treated as individual data
          samples for training a bout-level context classifier.}
        \label{fig:datarep}
\end{figure}

Once the full dataset of windows, with the corresponding contextual ground
truth labels, are extracted from the raw 2.5 hour data, further filtering is
applied to detect all those windows that do not contain any gait-related
activity. This is achieved by analysing individual windows and extracting their
DMOs. Mobility descriptors can be extracted from the vertical component of the
raw inertial signal sampled by the IMU sensor. For this task, we adopted a
state-of-the-art tool named Gaitpy \cite{DBLP:journals/jossw/CzechP19}. Gaitpy
is an open-source package that allows both the detection of walking bouts
within an inertial signal, and the extraction of a number of DMOs from regions
of interest. Specifically, Gaitpy works in a 2-stage process. It firsts chunks
the raw vertical acceleration signal into non-overlapping windows of 3 seconds
of duration, or \textit{epochs}, that are classified as gait or non-gait using
pre-trained machine learning models. After that, the generated predictions are
used to detect walking bouts, if any, and DMOs are computed. Gaitpy isolates
walking cycles for every detected bout, and returns the following features:
number of steps, step duration and step duration asymmetry, step length and
step length asymmetry, stride length and stride length asymmetry, stride
duration and stride duration asymmetry, cadence, initial and terminal double
support with asymmetry, single limb support and single limb support asymmetry,
stance and stance asymmetry, swing and swing asymmetry, gait speed. An example of a window
containing different walking bouts is shown in Figure \ref{fig:window}, where
Gaitpy correctly isolates the 3 different gait events highlighted in the red
boxes. With this setup, 2 different groups of datasets can be produced from the
raw acceleration signals.

\begin{itemize}
\item \textbf{Tabular DMO datasets}: DMOs extracted with Gaitpy always relate
  to individual walking bouts. As most windows contain more than one walking
  bout, DMOs can either be collected at a walking bout level or aggregated for
  every window. DMO aggregation is performed by summing the number of step for
  all the bouts in every window, and averaging the rest of the descriptors.
\item \textbf{Inertial signal datasets}: these 2 datasets contain the raw
  inertial signals sampled from the IMUs, where each data point is either a
  full 1 minute window, or a walking bout extracted using the temporal
  coordinates produced by Gaitpy (as shown in Figure \ref{fig:bouts}).
\end{itemize}


Table \ref{table:windows} contains a compendium of the datasets described
above. Data are obtained from a total of 9 participants, recruited across 3
different research centres. All participants are healthy adults, hence they are not
affected by any conditions that may impact the quality or quantity of their
activities. A total of 1075 windows were generated, distributed as 805 indoor
windows and 270 outdoor windows. This includes all the windows from the 2.5
hour data for which a contextual label was available. Windows for which DMOs
were available sum up to 311, distributed as 241 indoor windows and 70 outdoor
windows. These windows are a subset of the 1075 initial windows. A total of 434
walking bouts is extracted from the 311 DMO windows, distributed as 340 indoor
bouts and 94 outdoor bouts. Walking bouts have variable duration, bounded to a
maximum of 60 seconds (that is, the duration of the enclosing window).


\begin{table*}
\centering
\caption{\label{table:windows}Windows and walking bouts distribution across target
  participants and context classes.}
\begin{tabular}{r|rrr|rrr|rrr}
  subject & wins. & indoor     & outdoor    & DMO wins. & indoor     & outdoor   & bouts & indoor & outdoor \\\hline
  1       & 183   & 176        & 7          & 64        & 64         &  0        & 91  & 91  &  0 \\
  2       & 64    & 33         & 31         & 11        &  4         &  7        & 14  &  4  & 10 \\
  3       & 194   & 0          & 194        & 52        &  0         & 52        & 70  &  0  & 70 \\
  4       & 4     & 0          & 4          &  1        &  0         &  1        &  1  &  0  &  1 \\
  5       & 179   & 179        & 0          & 65        & 65         &  0        & 89  & 89  &  0 \\
  6       & 69    & 69         & 0          & 26        & 26         &  0        & 42  & 42  &  0 \\
  7       & 177   & 176        & 1          & 55        & 55         &  0        & 80  & 80  &  0 \\
  8       & 180   & 147        & 33         & 32        & 22         & 10        & 42  & 29  & 13 \\
  9       & 25    & 25         & 0          &  5        &  5         &  0        &  5  &  5  &  0 \\\hline
          & 1075  & 805 (.749) & 270 (.251) & 311       & 241 (.775) & 70 (.225) & 434 & 340 (.783) & 94 (.217)
\end{tabular}
\end{table*}


%% file: chapters/4.experiments.tex
\section{Experimental Setup}\label{section:experimentalcampaigns}
The primary focus of this study is to compare the classification accuracy of
machine learning algorithms which use raw inertial signals as input data,
against those that use gait descriptors as explicitly extracted domain
features, for the task of context detection. For this purpose, we design and
run 4 different experimental campaigns, obtained as combinations of the target
datasets and the classification techniques. A global overview of the conducted 
experiments is outlined in Table \ref{table:experiments}.

\begin{table*}
\centering
\caption{Experimental campaigns arranged by dataset and classification approach}
\label{table:experiments}
\begin{tabular}{l|ll|}
\cline{2-3}
 & \multicolumn{1}{c}{DMO windows} & \multicolumn{1}{c|}{walking
  bouts}\\ \hline

\multicolumn{1}{|l|}{DMOs} & \begin{tabular}[c]{@{}l@{}} All windows annotated
  with an indoor/outdoor ground truth\\ label, for which DMOs can be
  extracted. As one window\\ may contain multiple walking bouts, further
  decomposed into\\ walking cycles, features are aggregated on a gait cycle
  level,\\ and later normalized using:\\ \\ - z-score normalization\\ - no
  normalization\\\\Results in Table \ref{table:dmowindowsDMO}.\end{tabular}
& \begin{tabular}[c]{@{}l@{}} Individual walking bouts extracted from the
    windows, labeled\\ with indoor/outdoor ground truth labels. Features are
    extracted\\ for each individual walking bout, and later normalized
    using:\\ \\ - z-score normalization\\ - no normalization\\\\Results in
    Table \ref{table:walkingboutsDMO}.\end{tabular} \\ \hline

\multicolumn{1}{|l|}{raw signal} & \begin{tabular}[c]{@{}l@{}} All windows
  annotated with an indoor/outdoor ground truth\\ label, for which DMOs can be
  extracted. TSC methods can be\\ trained using one of the following
  signals:\\ \\ - vertical acceleration\\ - magnitude\\\\Results in Table
  \ref{table:dmowindowsTSC}.\end{tabular} & \begin{tabular}[c]{@{}l@{}}
  Individual walking bouts extracted from the windows annotated\\ with
  indoor/outdoor ground truth labels. TSC methods can be\\ trained using one of
  the following signals:\\ \\ - vertical acceleration\\ -
  magnitude\\ \\ Possible approaches to series of different length are
  padding,\\ length resampling, or keeping the original length when
  possible.\\\\Results in Table \ref{table:walkingboutsTSC}.\end{tabular}
\\ \hline

\end{tabular}
\end{table*}

\subsection{Raw Signal Classification}\label{subsection:tscmethods}
Windows and walking bouts can be classified using the IMU raw signals, or any
other time-dependant quantity derived from acceleration and angular
velocity. Leveraging on inertial signals for context detection allows to train
classification models that do not require overly engineered feature sets, and
can be tuned by tweaking few hyperparameters.
For this experimental campaign, we decided to focus on end-to-end univariate time series
classification methods. Hence, two different signals are separately extracted and evaluated:

\begin{itemize}
    \item the vertical component of the acceleration (which is the same
      acceleration component used by Gaitpy to extract DMOs)
    \item the acceleration magnitude, computed as: $$mag = \sqrt{acc_x^2 +
      acc_y^2 + acc_z^2}$$
\end{itemize}

Whilst windows always have the same length of one minute, corresponding to 6000
data points sampled at 100 Hz, walking bouts may differ in duration. As some of
the evaluated classification methods are not able to work with variable-length
data \cite{dempster_etal_2020_rocket},
\cite{DBLP:journals/corr/abs-2006-01667}, walking bouts shorter than 6000
points can either be padded with leading and trailing zeros until they all
reach the same length, or upsampled to the required length (see Figure
\ref{fig:datarep}). Walking bouts of different length, thus respecting their
original time amplitude, can be handled by MrSEQL.

The target time series models evaluated here, and presented in Section
\ref{section:related}, are ROCKET and its variant MiniROCKET, the strong
baseline 1nn-DTW, the symbolic linear classifiers WEASEL, and MrSEQL. The popular 
Python \textit{sktime} library  \cite{DBLP:journals/corr/abs-1909-07872} was used
for their practical implementation. The configuration setup and parameters for
each algorithm are highlighted in Table \ref{table:params}. It is worth
mentioning that most of the parameters were left configured to their default
values as recommended in the original studies which proposed these algorithms,
as the target of the study is the comparison of TSC techniques against
classification based on gait descriptors (i.e., feature-based models). A classical stratified 5-fold
cross-validation protocol is used for performance evaluation.

\subsection{DMO-based Classification}\label{subsection:dmoclassification}

\begin{table}
\centering
\caption{Parameter settings for the classification models. Double line separates the TSC models from the classic feature-based models.}\label{table:params}
\begin{tabular}{ll}
\hline
\multicolumn{1}{c}{Algorithm} & \multicolumn{1}{c}{training configuration}\\ \hline
ROCKET                    & \begin{tabular}[c]{@{}l@{}}-
10000 kernels used (20000 features extracted)\end{tabular} 
\\ \hline
MiniROCKET                & \begin{tabular}[c]{@{}l@{}}- 10000 features extracted\\ - 32 maximum dilation values per kernel\end{tabular}                                                                   \\ \hline
1-nn DTW                  & \begin{tabular}[c]{@{}l@{}}- single neighbor used for best distance\\ - DTW as distance measure\end{tabular}                                                                   \\ \hline
WEASEL                    & \begin{tabular}[c]{@{}l@{}}- SFA bigrams included\\ - information gain used as binning strategy for\\~~breakpoint derivation\end{tabular}                                       \\ \hline
MrSEQL                    & \begin{tabular}[c]{@{}l@{}}- SAX used for symbolic representation\\ - SEQL mode is set as feature selection \end{tabular}                      \\ \hline\hline
Logistic Regression & \begin{tabular}[c]{@{}l@{}}- L2 regularization\\ - class weights balanced on frequency\\ - 1000 maximum iterations\end{tabular} \\\hline
Random Forest       & \begin{tabular}[c]{@{}l@{}}- 500 estimators\\ - Gini index as split criterion\end{tabular} \\\hline
KNN                 & \begin{tabular}[c]{@{}l@{}}- 5 neighbors\\ - uniform neighbor weights\end{tabular} \\\hline
SVM                 & - RBF kernels \\\hline
Ridge Regression    & \begin{tabular}[c]{@{}l@{}}- alphas equally spaced in logarithmic scale\\ - normalized regressors\end{tabular} \\\hline
\end{tabular}
\end{table}

As illustrated in Section \ref{subsection:datasets}, a number of DMOs can be
extracted from signal windows where gait-related activity is present. DMOs can
then be used as input features to train feature-based learning models. For our experimental campaign, we adopt
6 common classifiers, namely, Logistic Regression (Logistic), Random Forest
(RndForest), K-nearest neighbors (KNN), Support Vector Machine (SVM), Gaussian
Naive Bayes (GNB), and Ridge Regression (Ridge). Table \ref{table:params}
contains the parameters that are used to configure these models during the
experiments. Similarly to the TSC campaign, DMO classifiers are evaluated using
a stratified 5-fold cross-validation policy.

%% file: chapters/5.results.tex
\section{Results}\label{section:results}

The classification results for context detection of activity windows, are
outlined in Table \ref{table:dmowindowsDMO} and Table
\ref{table:dmowindowsTSC}, for the feature-based methods and the TSC methods
respectively. Reported metrics include accuracy, precision, recall and
F1-score. ROCKET and MrSEQL achieved the best classification accuracy ($> 96\%$), with
reasonable precision and recall scores that suggest no overfitting towards the
indoor majority class. Feature-based models did not show the same behaviour, as
SVM presents the highest accuracy value of 77.4\% (in the case of
non-normalized features) paired with a F1-score of 43.6\%. All these models,
with the exception of the Logistic Classifier, fall within the same accuracy
range, however, GNB shows the best precision and recall scores across the
board. Feature normalization does not seem to have a significant impact on the
models. Logistic Regression shows an accuracy and F1 improvement of 4\% when
normalized features are used, and KNN and SVM display the opposite behaviour for accuracy, but similar fro F1.
Random Forest, GNB and Ridge Regression were not affected by the feature scale. In a
similar fashion, the choice of signal between the vertical acceleration and the
magnitude for the TSC classifiers results in a marginal performance difference.
The main observation from these experiments is that there is a very significant gap between the feature-based models (F1: 61.3\%) and the TSC models (F1: 98.0\%).

\begin{table*}
  \centering
  \caption{Feature-based models on gait windows dataset}
  \label{table:dmowindowsDMO}
  \begin{tabular}{ll||llll}
    \hline
    model     & norm.   & accuracy       & precision      & recall         & F1-score       \\ \hline
    Logistic      & z-score & 66.2 $\pm$ 3.6 & 60.6 $\pm$ 4.2 & 64.5 $\pm$ 6.3 & 60.0 $\pm$ 4.3 \\
    Logistic      & -       & 62.0 $\pm$ 3.8 & 58.0 $\pm$ 4.8 & 61.3 $\pm$ 6.9 & 56.6 $\pm$ 4.8 \\ \hline
    RndForest        & z-score & 76.8 $\pm$ 2.2 & 65.3 $\pm$ 7.6 & 56.1 $\pm$ 2.1 & 56.1 $\pm$ 2.7 \\
    RndForest        & -       & 76.8 $\pm$ 2.2 & 65.3 $\pm$ 7.6 & 56.1 $\pm$ 2.1 & 56.1 $\pm$ 2.7 \\ \hline
    KNN       & z-score & 75.8 $\pm$ 3.1 & 62.1 $\pm$ 6.5 & 57.0 $\pm$ 3.7 & 57.6 $\pm$ 4.5 \\
    KNN       & -       & 76.2 $\pm$ 5.9 & 64.8 $\pm$ 14. & 55.2 $\pm$ 6.0 & 55.5 $\pm$ 7.1 \\ \hline
    SVM       & z-score & 76.5 $\pm$ 1.5 & 38.6 $\pm$ 0.2 & 49.3 $\pm$ 0.9 & 43.3 $\pm$ 0.4 \\
    SVM       & -       & 77.4 $\pm$ 0.1 & 38.7 $\pm$ 0.0 & 50.0 $\pm$ 0.0 & 43.6 $\pm$ 0.0 \\ \hline
    GNB       & z-score & 72.3 $\pm$ 9.3 & 63.1 $\pm$ 6.7 & 62.3 $\pm$ 6.2 & \textbf{61.3} $\pm$ \textbf{7.5} \\
    GNB       & -       & 72.3 $\pm$ 9.3 & 63.1 $\pm$ 6.7 & 62.3 $\pm$ 6.2 & \textbf{61.3} $\pm$ \textbf{7.5} \\ \hline
    Ridge     & z-score & 76.2 $\pm$ 1.6 & 38.5 $\pm$ 0.1 & 49.1 $\pm$ 1.1 & 43.2 $\pm$ 0.5 \\
    Ridge     & -       & 76.2 $\pm$ 1.6 & 38.5 $\pm$ 0.1 & 49.1 $\pm$ 1.1 & 43.2 $\pm$ 0.5 \\ \hline
  \end{tabular}
\end{table*}

\begin{table*}
  \centering
  \caption{TSC models on gait windows dataset}
  \label{table:dmowindowsTSC}
  \begin{tabular}{ll||llll}
    \hline
    model      & signal  & accuracy       & precision      & recall         & F1-score       \\ \hline
    ROCKET     & v. acc. & 98.7 $\pm$ 2.8 & 98.5 $\pm$ 3.1 & 97.6 $\pm$ 5.2 & \textbf{98.0} $\pm$ \textbf{4.3} \\
    ROCKET     & mag.    & 98.4 $\pm$ 3.5 & 97.8 $\pm$ 4.7 & 97.4 $\pm$ 5.7 & \textbf{97.6} $\pm$ \textbf{5.2} \\ \hline
    miniROCKET & v. acc. & 89.3 $\pm$ 5.3 & 89.6 $\pm$ 8.3 & 79.4 $\pm$ 9.9 & 82.3 $\pm$ 9.5 \\
    miniROCKET & mag.    & 90.3 $\pm$ 3.8 & 88.5 $\pm$ 5.1 & 82.6 $\pm$ 6.9 & 84.8 $\pm$ 6.2 \\ \hline
    1-nn DTW   & v. acc. & 65.9 $\pm$ 5.1 & 42.8 $\pm$ 6.9 & 44.5 $\pm$ 5.1 & 43.1 $\pm$ 5.4 \\
    1-nn DTW   & mag.    & 69.4 $\pm$ 4.4 & 51.9 $\pm$ 6.5 & 51.3 $\pm$ 4.3 & 50.8 $\pm$ 4.8 \\ \hline
    WEASEL     & v. acc. & 79.4 $\pm$ 6.6 & 73.6 $\pm$ 10. & 70.8 $\pm$ 12. & 70.7 $\pm$ 13. \\
    WEASEL     & mag.    & 81.0 $\pm$ 4.6 & 79.1 $\pm$ 7.9 & 75.9 $\pm$ 8.1 & 75.2 $\pm$ 7.1 \\ \hline
    MrSEQL     & v. acc. & 96.1 $\pm$ 8.5 & 95.9 $\pm$ 9.1 & 93.6 $\pm$ 14. & 94.1 $\pm$ 13. \\
    MrSEQL     & mag.    & 95.5 $\pm$ 9.9 & 97.6 $\pm$ 5.2 & 91.7 $\pm$ 18. & 91.6 $\pm$ 18. \\ \hline
  \end{tabular}
\end{table*}

Table \ref{table:walkingboutsDMO} and Table \ref{table:walkingboutsTSC} show
the classification results on the walking bout dataset, for feature-based
methods and TSC methods respectively. The reported metrics for this dataset are
aligned with the window dataset results previously discussed, in terms of both
performance gap between the two method groups, and sensitivity to feature
scale and input signals. In the TSC group, walkin bouts are either padded or
resampled so that they all have the same length. Once again, MrSEQL and ROCKET
are the best performing models all-around, with the latter being slightly more
consistent in terms of precision and recall. Padding seems to be the ideal
strategy to manage bout length, while choosing vertical acceleration or
magnitude has a very limited impact on the model performance.

\begin{table*}
  \centering
  \caption{Feature-based models on walking bouts dataset}
  \label{table:walkingboutsDMO}
  \begin{tabular}{ll||llll}
    \hline
    & \multicolumn{4}{c}{walking bouts}                                 \\ \hline
    model     & norm.   & accuracy       & precision      & recall         & F1-score       \\ \hline
    Logistic  & z-score & 64.0 $\pm$ 4.7 & 58.4 $\pm$ 4.0 & 61.6 $\pm$ 5.5 & 57.4 $\pm$ 4.5 \\
    Logistic  & -       & 61.9 $\pm$ 4.6 & 58.7 $\pm$ 3.3 & 62.6 $\pm$ 4.9 & 56.8 $\pm$ 4.0 \\ \hline
    RndForest & z-score & 77.4 $\pm$ 1.0 & 62.5 $\pm$ 3.6 & 54.7 $\pm$ 2.5 & 54.0 $\pm$ 3.9 \\
    RndForest & -       & 77.4 $\pm$ 1.5 & 62.9 $\pm$ 2.9 & 54.8 $\pm$ 1.6 & 54.2 $\pm$ 3.1 \\ \hline
    KNN       & z-score & 75.3 $\pm$ 3.0 & 59.6 $\pm$ 6.8 & 56.1 $\pm$ 4.4 & 56.5 $\pm$ 5.2 \\
    KNN       & -       & 76.2 $\pm$ 2.9 & 59.5 $\pm$ 10. & 53.6 $\pm$ 3.5 & 53.0 $\pm$ 4.8 \\ \hline
    SVM       & z-score & 77.6 $\pm$ 1.0 & 45.8 $\pm$ 15. & 50.3 $\pm$ 2.4 & 45.4 $\pm$ 4.2 \\
    SVM       & -       & 78.3 $\pm$ 0.4 & 39.1 $\pm$ 0.2 & 50.0 $\pm$ 0.0 & 43.9 $\pm$ 0.1 \\ \hline
    GNB       & z-score & 75.5 $\pm$ 3.8 & 64.8 $\pm$ 4.6 & 65.5 $\pm$ 5.2 & \textbf{64.9} $\pm$ \textbf{4.8} \\
    GNB       & -       & 75.5 $\pm$ 3.8 & 64.8 $\pm$ 4.6 & 65.5 $\pm$ 5.2 & \textbf{64.9} $\pm$ \textbf{4.8} \\ \hline
    Ridge     & z-score & 76.9 $\pm$ 1.7 & 42.4 $\pm$ 7.5 & 49.4 $\pm$ 1.3 & 44.3 $\pm$ 2.0 \\
    Ridge     & -       & 76.9 $\pm$ 1.7 & 42.4 $\pm$ 7.5 & 49.4 $\pm$ 1.3 & 44.3 $\pm$ 2.0 \\ \hline
  \end{tabular}
\end{table*}

\begin{table*}
  \centering
  \caption{TSC models on walking bouts dataset}
  \label{table:walkingboutsTSC}
  \begin{tabular}{lll||llll}
    \hline
    model      & series lenght & signal    & accuracy       & precision      & recall         & F1-score       \\ \hline
    ROCKET     & pad           & v. acc.   & 94.7 $\pm$ 3.6 & 95.5 $\pm$ 4.8 & 88.5 $\pm$ 6.8 & 91.4 $\pm$ 6.2 \\
    ROCKET     & pad           & mag.      & 96.0 $\pm$ 3.6 & 96.4 $\pm$ 4.7 & 91.7 $\pm$ 6.7 & 93.7 $\pm$ 6.0 \\
    ROCKET     & resample      & v. acc.   & 92.8 $\pm$ 2.5 & 94.5 $\pm$ 2.7 & 84.2 $\pm$ 5.7 & 87.9 $\pm$ 4.9 \\
    ROCKET     & resample      & mag.      & 96.0 $\pm$ 3.2 & 97.0 $\pm$ 3.2 & 91.3 $\pm$ 6.6 & 93.6 $\pm$ 5.4 \\ \hline
    miniROCKET & pad           & v. acc.   & 87.3 $\pm$ 2.1 & 85.2 $\pm$ 4.3 & 74.9 $\pm$ 3.5 & 78.3 $\pm$ 3.6 \\
    miniROCKET & pad           & mag.      & 86.1 $\pm$ 1.7 & 83.8 $\pm$ 3.9 & 72.3 $\pm$ 2.3 & 75.8 $\pm$ 2.8 \\
    miniROCKET & resample      & v. acc.   & 90.5 $\pm$ 1.7 & 90.5 $\pm$ 3.6 & 81.1 $\pm$ 5.4 & 84.2 $\pm$ 3.8 \\
    miniROCKET & resample      & mag.      & 91.0 $\pm$ 2.4 & 90.7 $\pm$ 5.5 & 81.8 $\pm$ 3.8 & 85.2 $\pm$ 4.1 \\ \hline
    1-nn DTW   & pad           & v. acc.   & 59.9 $\pm$ 4.0 & 49.1 $\pm$ 2.1 & 49.0 $\pm$ 2.9 & 48.3 $\pm$ 2.4 \\
    1-nn DTW   & pad           & mag.      & 58.0 $\pm$ 3.0 & 49.4 $\pm$ 1.7 & 49.3 $\pm$ 2.4 & 48.2 $\pm$ 1.8 \\
    1-nn DTW   & resample      & v. acc.   & 46.1 $\pm$ 5.9 & 55.2 $\pm$ 5.1 & 56.3 $\pm$ 6.6 & 44.5 $\pm$ 5.1 \\
    1-nn DTW   & resample      & mag.      & 43.7 $\pm$ 5.7 & 55.7 $\pm$ 4.4 & 56.8 $\pm$ 5.5 & 43.0 $\pm$ 5.3 \\ \hline
    WEASEL     & pad           & v. acc.   & 82.9 $\pm$ 7.4 & 83.8 $\pm$ 11. & 70.2 $\pm$ 5.4 & 71.9 $\pm$ 6.8 \\
    WEASEL     & pad           & mag.      & 83.4 $\pm$ 6.2 & 81.5 $\pm$ 11. & 70   $\pm$ 4.4 & 72.5 $\pm$ 5.8 \\
    WEASEL     & resample      & v. acc.   & 81.3 $\pm$ 6.5 & 73.8 $\pm$ 7.2 & 72.1 $\pm$ 6.8 & 72.3 $\pm$ 7.5 \\
    WEASEL     & resample      & mag.      & 84.3 $\pm$ 2.9 & 78.3 $\pm$ 5.1 & 72.9 $\pm$ 7.8 & 74   $\pm$ 7.6 \\ \hline
    MrSEQL     & pad           & v. acc.   & 97.7 $\pm$ 5.1 & 98.7 $\pm$ 2.8 & 94.7 $\pm$ 11. & \textbf{95.7} $\pm$ \textbf{9.5} \\
    MrSEQL     & pad           & mag.      & 97.4 $\pm$ 5   & 98.5 $\pm$ 2.8 & 94.2 $\pm$ 11. & \textbf{95.4} $\pm$ \textbf{9.3} \\
    MrSEQL     & resample      & v. acc.   & 95.8 $\pm$ 9.2 & 97.9 $\pm$ 4.6 & 90.5 $\pm$ 21. & 89.8 $\pm$ 22. \\
    MrSEQL     & resample      & mag.      & 95.6 $\pm$ 9.7 & 92.8 $\pm$ 15. & 90.3 $\pm$ 21. & 89.7 $\pm$ 23. \\
    MrSEQL     & original      & v. acc.   & 89.4 $\pm$ 2.6 & 94.1 $\pm$ 1.3 & 75.6 $\pm$ 6   & 80.2 $\pm$ 6.2 \\
    MrSEQL     & original      & mag.      & 90.8 $\pm$ 2 & 93.9 $\pm$ 2.1 & 79.1 $\pm$ 4.6 & 83.6 $\pm$ 4.2 \\ \hline
  \end{tabular}
\end{table*}


%% file: chapters/6.discussion.tex
\section{Discussion}\label{section:discussion}

The ultimate goal of this paper is to understand whether inertial signals
alone, sampled during gait-related activities, suffice for the task of
indoor/outdoor context detection, retaining enough information so that GPS data
are not needed. In relation to this first research question, results of the
experiments on both the gait windows dataset and the walking bouts dataset, indicate
that context can indeed be inferred from inertial movement descriptors, either using 
DMO-features extracted from the raw signals or the raw signals themselves, without
further inclusion of geographical coordinates. This suggests that the
contextual environment in which gait episodes occur, has an impact on the
subjects' walking quality, most likely caused by the difference in terrain,
footwear, weather conditions, and motion utility.

Our secondary objective was to understand how classical feature-based methods,
trained with DMOs extracted from the raw signals, compare with TSC methods on
the context detection task. The most interesting finding to emerge from the
experimental results in Section \ref{section:results} is that time series
models greatly outperform feature-based models, regardless of the target metric
and dataset. TSC learners achieve high accuracy and are proven to be more versatile as they can be used to classify any
type of signal. In fact, many of the labeled windows extracted from the data
 had to be discarded as DMOs could not be extracted (no gait detected by Gaitpy), but they can still be classified if treated purely as time series. The performance gap
between the two types of methods can be explained in part by the fact that DMOs represent a limited set of shallow features, whilst most TSC algorithms
are based on data transformation steps that result into rich, context-based descriptors. This allows the identification of hidden or underlying patterns in
the signals that may otherwise go unnoticed.

\subsection{Limitations}
The results obtained during this work are encouraging and seem to suggest that context can effectively be inferred from inertial signals sampled during
walking activities. However, a number of limiting aspects of this work should
be further discussed, that may prompt interesting starting points for future
studies.

\subsubsection*{Class Imbalance and Validation}
The target datasets used for this study suffer from a severe class imbalance,
as the majority indoor class constitutes 75-78\% of the total data points,
depending on the dataset. Moreover, the class imbalance is even more severe
across subjects, as only 3 participants out of 9 contribute with a significant
amount of outdoor activities. This poses a non-trivial limitation to the
validation protocols that can be used. The most common choice in this sense, is
the Leave-One-Subject-Out (LOSO) cross-validation protocol
\cite{DBLP:journals/corr/abs-1806-05226}, where folds for the cross-validation
process are designed based on the participants in the dataset. This ensures
that each participant is independently used as validation set only once, and no
positive bias is introduced in the models. In order to ensure that our
experimental results are not affected by subject overfitting, we conducted 3
separate training campaigns for our best-performing models: one based on fully
fledged LOSO cross-validation (9 splits), and two based on custom splits where
subjects do not simultaneously appear in the train set and in test set. Custom
splits were designed prior to the training process. In all these campaigns, the
performance of the models did not deteriorate significantly in terms of
classification accuracy and F1-score, while standard deviation for all the
registered metrics marginally increased. This suggests that the results
presented in Section \ref{section:results} are not obtained by models that
merely learned the target subjects rather than the underlying patterns of the
sampled walking activities.

\subsubsection*{Interpretability}
Context inference for gait-related activities is set to provide insights on
human movement that may potentially be used as diagnostic and clinical
indicators. Therefore, it is paramount for models that compute said insights to
be interpretable, and to offer simple explanations on how activities are
classified.

In this regard, feature-based classifiers that leverage activity
DMOs are easily interpretable, as mobility outcomes are simple descriptors that
measure natural gait quantities, such as number of steps, cadence, or
stride. In order to better understand how DMOs impact the final decisions of
our feature-based models, we evaluated the feature importance scores for the
Logistic Regression and the Ridge Regression models. Not surprisingly, only a
subset of DMOs have a significant impact on the final model predictions: while
in most cases step duration, step length, and swing asimmetry are primary
indicators of indoor walking, the number of steps, stride length, and swing are
mainly used to predict the outdoor class.

Whilst tabular models have straightforward and intuitive interpretation
techniques such as feature importance scores, explainability for TSC methods is
a broader and more novel topic, that is recently gaining traction and attention
in the context of Explainable Artificial Intelligence
\cite{DBLP:journals/corr/abs-2104-04075}. The most straightforward approach to
interpret the decisions made by a TSC model is to use \textit{saliency maps}
\cite{10.1007/978-3-030-65742-0_6}, a visual tool that highlights important
regions within a target time series, that significantly contribute to the final
model predictions. In contrast with feature importance, saliency maps rely on
vectors of weights that are assigned to every point in a given series. In this
study, we briefly tested MrSEQL-SM \cite{DBLP:journals/corr/abs-2006-01667}
using the MrSEQL trained models. Detailed results on interpretation are out of the scope of this study, but we find this a worthile topic for  further study  in future work.


%% file: chapters/7.conclusions.tex
\section{Conclusions}\label{section:conclusions}
In this paper, we proposed a novel methodology for automated context detection
from raw inertial signals, as an effective tool for describing human mobility
data. We also presented a detailed empirical study concerned with the
application of recent state-of-the-art TSC algorithms to the problem of context
detection, and we compared their performance with more classical feature-based models
relying on extracted mobility descriptors (DMOs). Our experiments were based on inertial data collected
from 9 healthy individuals, annotated with metadata representing contextual
ground-truth labels. Both TSC methods and DMO-based methods were tested, and ultimately, we showed that automated context detection can effectively be achieved
with very high accuracy using only raw inertial signals, without the need for complex gait descriptor features.

While we obtained very promising results, we consider this study to be a detailed proof
of concept, set to provide the basis for a more extensive work. Next, we aim to expand
the cohort of target participants. This includes both the number of subjects involved in the
data collection process, and the clinical conditions subjects may be affected from. In fact,
within the Mobilise-D consortium, plenty of mobility data were collected from patients affected
by Parkinson's disease, multiple sclerosis, chronic obstructive pulmonary disease, and
recovering from proximal femoral fracture.

%% file: chapters/8.acknowledgements.tex
\section*{Acknowledgment}

This work was supported by the Mobilise-D project that has received funding
from the Innovative Medicines Initiative 2 Joint Undertaking (JU) under grant
agreement No. 820820. This JU receives support from the European Union's
Horizon 2020 research and innovation program and the European Federation of
Pharmaceutical Industries and Associations (EFPIA). Content in this publication
reflects the authors’ view and neither IMI nor the European Union, EFPIA, or
any Associated Partners are responsible for any use that may be made of the
information contained herein. This work was also supported by a grant from
Science Foundation Ireland through the Insight Centre for Data Analytics
(12/RC/2289\_P2). For the purpose of Open Access, the authors have applied a CC
BY public copyright licence to any Author Accepted Manuscript version arising
from this submission.


%% file: chapters/9.appendix.tex
\appendix
As extensively discussed in Section \ref{subsection:datasets}, all the datsets
in this study were generated from inertial signals containing gait
activity. However, after windowing the raw signals and filtering the windows
according to whether they include walking bouts or not, only 311 windows out of
1075 are left (see Table \ref{table:windows}). This is necessary in order to
make a direct comparison between feature-based methods and TSC methods. We
further validated our experimental results by classifying the full window
dataset, using TSC methods only. Results are presented in
Table \ref{table:fulldataraw}.

\begin{table*}[]
  \centering
  \caption{TSC classification results on full dataset}
  \label{table:fulldataraw}
  \begin{tabular}{ll||cccc}
    \hline
    model    & signal  & accuracy       & precision      & recall         & f1-score        \\ \hline
    ROCKET   & v. acc. & 93.6 $\pm$ 5.3 & 83.1 $\pm$ 9.5 & 86.4 $\pm$ 9.5 & 86.4 $\pm$ 9.5  \\
    ROCKET   & mag.    & 90.5 $\pm$ 6.6 & 75.4 $\pm$ 8.0 & 79.1 $\pm$ 9.1 & 79.1 $\pm$ 9.1  \\ \hline
    mROCKET  & v. acc. & 88.4 $\pm$ 2.4 & 88.7 $\pm$ 4.6 & 79.6 $\pm$ 3.1 & 82.8 $\pm$ 3.5  \\
    mROCKET  & mag.    & 89.3 $\pm$ 2.1 & 89.4 $\pm$ 3.0 & 81.4 $\pm$ 3.6 & 84.3 $\pm$ 3.3  \\ \hline
    1-nn DTW & v. acc. & 73.5 $\pm$ 1.6 & 65.1 $\pm$ 2.2 & 65.3 $\pm$ 2.8 & 65.1 $\pm$ 2.4  \\
    1-nn DTW & mag.    & 78.1 $\pm$ 1.8 & 70.6 $\pm$ 2.8 & 68.5 $\pm$ 2.6 & 69.3 $\pm$ 2.7  \\ \hline
    WEASEL   & v. acc. & 54.5 $\pm$ 25. & 64.9 $\pm$ 12. & 64.5 $\pm$ 11. & 51.2 $\pm$ 24.  \\
    WEASEL   & mag.    & 78.8 $\pm$ 7.0 & 74.6 $\pm$ 7.6 & 76.9 $\pm$ 4.7 & 74.5 $\pm$ 6.4  \\ \hline
    MrSEQL   & v. acc. & 95.6 $\pm$ 9.5 & 96.6 $\pm$ 7.0 & 91.5 $\pm$ 18. & 91.4 $\pm$ 18.  \\
    MrSEQL   & mag.    & 95.1 $\pm$ 10. & 93.8 $\pm$ 13. & 90.7 $\pm$ 20. & 90.1 $\pm$ 21.  \\ \hline
  \end{tabular}
\end{table*}

Table \ref{table:customsplits} shows the results obtained on a set of custom
validation splits, manually designed to ensure that our default 5-fold
cross-validation policy does not introduce positive bias due to the inclusion, for each
split, of data from the same participants in both the train set and the test set. With respect to
the subject IDs indicated in Table \ref{table:windows}, tested splits include 1) classical
LOSO, resulting in 9 total splits, 2) 3-fold splitting, where the 3 test sets are composed with
data from subject 2, subjects 3 and 5, and subject 8 respectively, and 3) 5-fold
splitting, where the test sets are composed with data from subjects 1, 2, and 8, subjects 3 and 5,
subjects 3 and 6, subjects 3 and 7, and subjects 4 and 9 respectively. For this experiment,
we solely focused on the fastest and the best-performing models, nameky, ROCKET,
miniROCKET, and the baseline 1-nn DTW. All custom splits were tested on the full window dataset, using
vertical acceleration and acceleration magnitude.

\begin{table*}[]
  \centering
  \caption{TSC classification results on full dataset, using custom splits}
  \label{table:customsplits}
  \begin{tabular}{lll||cccc}
    \hline
    model    & split   & split   & accuracy       & precision      & recall         & f1-score        \\ \hline
    ROCKET   & LOSO    & v. acc. & 93.5 $\pm$ 6.0 & 75.6 $\pm$ 22  & 73.2 $\pm$ 21  & 73.2 $\pm$ 21.  \\
    ROCKET   & LOSO    & mag.    & 86.1 $\pm$ 15. & 63.5 $\pm$ 20. & 55.5 $\pm$ 21. & 57.6 $\pm$ 19.  \\
    ROCKET   & 3-fold  & v. acc. & 84.7 $\pm$ 19. & 87.4 $\pm$ 15. & 83.0 $\pm$ 18. & 81.6 $\pm$ 22.  \\
    ROCKET   & 3-fold  & mag.    & 84.2 $\pm$ 18. & 88.2 $\pm$ 13. & 82.8 $\pm$ 18. & 80.7 $\pm$ 22.  \\
    ROCKET   & 5-fold  & v. acc. & 91.8 $\pm$ 15. & 89.7 $\pm$ 20. & 89.4 $\pm$ 20. & 89.1 $\pm$ 20.  \\
    ROCKET   & 5-fold  & mag.    & 91.2 $\pm$ 12. & 87.5 $\pm$ 19. & 87.9 $\pm$ 19. & 87.5 $\pm$ 19.  \\ \hline
    mROCKET  & LOSO    & v. acc. & 60.6 $\pm$ 35. & 53.2 $\pm$ 25. & 45.5 $\pm$ 30. & 42.2 $\pm$ 27.  \\
    mROCKET  & LOSO    & mag.    & 61.8 $\pm$ 30. & 51.5 $\pm$ 23. & 42.4 $\pm$ 27. & 44.4 $\pm$ 25.  \\
    mROCKET  & 3-fold  & v. acc. & 53.3 $\pm$ 3.9 & 49.2 $\pm$ 21. & 49.8 $\pm$ 4.0 & 40.1 $\pm$ 5.5  \\
    mROCKET  & 3-fold  & mag.    & 60.2 $\pm$ 9.9 & 63.0 $\pm$ 8.6 & 53.9 $\pm$ 3.3 & 47.4 $\pm$ 9.8  \\
    mROCKET  & 5-fold  & v. acc. & 52.6 $\pm$ 19. & 37.8 $\pm$ 8.3 & 48.1 $\pm$ 2.4 & 35.2 $\pm$ 8.8  \\
    mROCKET  & 5-fold  & mag.    & 55.2 $\pm$ 17. & 62.3 $\pm$ 9.8 & 52.3 $\pm$ 1.4 & 40.5 $\pm$ 11.  \\ \hline
    1-nn DTW & LOSO    & v. acc. & 54.3 $\pm$ 27. & 44.5 $\pm$ 16. & 38.4 $\pm$ 23. & 36.0 $\pm$ 16.  \\
    1-nn DTW & LOSO    & mag.    & 57.4 $\pm$ 30. & 45.7 $\pm$ 16. & 39.3 $\pm$ 24. & 37.3 $\pm$ 17.  \\
    1-nn DTW & 3-fold  & v. acc. & 51.1 $\pm$ 3.0 & 52.1 $\pm$ 8.3 & 52.2 $\pm$ 4.6 & 44.0 $\pm$ 4.3  \\
    1-nn DTW & 3-fold  & mag.    & 64.8 $\pm$ 8.0 & 62.7 $\pm$ 9.0 & 57.7 $\pm$ 3.4 & 55.5 $\pm$ 1.5  \\
    1-nn DTW & 5-fold  & v. acc. & 51.5 $\pm$ 10. & 52.0 $\pm$ 9.8 & 47.4 $\pm$ 6.8 & 41.2 $\pm$ 4.6  \\
    1-nn DTW & 5-fold  & mag.    & 51.4 $\pm$ 17. & 55.4 $\pm$ 13. & 49.1 $\pm$ 17. & 45.4 $\pm$ 13.  \\ \hline
  \end{tabular}
\end{table*}
